\documentstyle[preprint,aps]{revtex}

\tighten
\begin{document}
\preprint{Alberta-Thy-03-96}
\draft

\title{On the Stability of Spherical Membranes in Curved Spacetimes}
\author{A. L. LARSEN\thanks{Electronic Address:
alarsen@phys.ualberta.ca}}
\address{Theoretical Physics Institute, Department of Physics,\\
University of Alberta, Edmonton, Canada T6G 2J1}
\author{C. O. LOUSTO\thanks{Electronic Address:
lousto@mail.physics.utah.edu}}
\address{Department of Physics, University of Utah,\\
201 JFB,
Salt Lake City, UT 84112, USA}
\date{\today}
\maketitle
\begin{abstract}
\baselineskip=1.5em
We study the existence and stability of spherical membranes in
curved spacetimes. For Dirac membranes in the Schwarzschild--de
Sitter background we find that there exists an equilibrium
solution. By fine--tuning the dimensionless parameter $\Lambda M^2,$
the static
membrane can be at any position outside the black hole event horizon,
even at
the stretched horizon, but the solution is unstable. We show that
modes having
$l=0$ (and for $\Lambda M^2<16/243$ also $l=1$) are responsible
for the instability. We also find that spherical higher order
membranes (membranes with extrinsic curvature corrections), contrary
to what
happens in flat Minkowski space, {\it do} have
equilibrium solutions in a general curved background and, in
particular, also in the ``plain'' Schwarzschild geometry (while Dirac
membranes do not have equilibrium solutions there). These solutions,
however,
are also unstable.
We shall discuss a way of by--passing these instability problems,
and we also relate our results to the recent ideas of representing
the black
hole event horizon as a relativistic bosonic membrane.

\end{abstract}


\section{Introduction}
\setcounter{equation}{0}

The idea of representing physical objects by extended ones
such as membranes, can be traced back to the pioneering work
of Dirac\ \cite{D62}. In this model for the electron, tension
forces tending to collapse the membrane were balanced by
repulsive electric forces. Later, canonical\ \cite{CT76,AS90}
and semiclassical\ \cite{DIPSS88} quantization of membranes in
a flat background was considered. On the other hand, starting
with the works of `t Hooft\ \cite{tH85}, a series of
papers\cite{STU93,M94,L295} considered
the quantum degrees of freedom of a black hole, localized around
its event horizon. It is also worth noting the classical approach
to black holes by the, so called, ``membrane paradigm'',
summarized in Ref.\ \cite{TPM86}.

In Ref.\ \cite{L95} it was considered the approach of
effectively representing
the quantum degrees of freedom of a black hole by a relativistic
membrane. The total Lagrangian of the system was composed of the
classical curved geometry plus a relativistic bosonic membrane.
The membrane, located at the ``stretched horizon'', was decomposed
into a spherically symmetric term plus small fluctuations around it.
The first order fluctuations have then been quantized and the energy
spectrum was obtained. From this, the entropy of the membrane at the
temperature of the black hole (given by the background geometry) was
computed. It was however left open the question of the existence and
stability of the zeroth order solution representing a spherical
membrane at rest very close to the black hole event horizon.
In the present paper we address the above question.

This paper is organized as follows: In Sec.\ II we study the
Dirac membrane in its simplest form. No coupling to the
electromagnetic field is considered, instead the membrane
is let to freely move in a curved background. Since
our final aim is to study black hole properties, and for the
sake of simplicity, we consider spherically symmetric
backgrounds. We derive the condition for the existence of
spherical membranes to be in equilibrium at a given radius
$r_m$. We also find the stability condition and apply it to
a membrane in the Schwarzschild--de Sitter space. Sec.\ III
deals with a covariant description of small fluctuations around the  
equilibrium
position $r_m$. We find that the fluctuations are governed by a
Klein--Gordon--like equation in the 2+1 dimensional
world--volume swept by
the membrane. We identify modes with angular momentum
label $l=0,1$ as responsible for the instabilities in the
Schwarzschild--de Sitter background. In Sec.\ IV we study
a more general membrane described by a Lagrangian with up to
quadratic terms in the extrinsic curvature. This introduces
two more arbitrary constants $A$ and $B$, in addition to the
membrane tension $T$. This freedom allows us, in the plain  
Schwarzschild
background,  to choose $r_m$
as close to the event horizon as we want
(for instance at the stretched horizon). We
study the radial small fluctuation equation (fourth order  
differential
equation)
and find the equilibrium and stability conditions. In
Sec.\ V we discuss the way of extending the validity of
the perturbative approach around a maximum of the potential,
instead of a minimum, when we take the comoving system of
reference to describe the zeroth order motion. In
Appendix\ A we give the analytic expressions, corresponding
to the plots of Fig.\ \ref{fig1}, for the cosmological and
event horizon, and $r_m$ in the Schwarzschild--de Sitter
spacetime.

\section{The Spherical Dirac Membrane}
\setcounter{equation}{0}

Let us consider the action of a Dirac membrane (with tension
$T$ bearing dimension of $length^{-3}$) in a curved background
\begin{equation}
I_{M}=-T\int d\tau d\rho d\sigma  \sqrt{-{\mbox{det}}
\left(\gamma_{ij}\right)}~,
\end{equation}
where $\gamma_{ij}$ is the induced metric on the world-volume
\begin{equation}
\gamma_{ij}=g_{\mu\nu}\partial_i x^{\mu}\partial_j x^{\nu}~.
\label{1.1}
\end{equation}
The classical equations of motion derived from this action are
\begin{equation}
\Box_\gamma{x^{\mu}}+\gamma^{ij}
\Gamma^\mu_{\kappa\lambda}(g)\partial_i{x^{\kappa}}
\partial_j{x^{\lambda}}=0~,
\label{1.2}
\end{equation}
where
\begin{equation}
\Box_\gamma={1\over\sqrt{-\gamma}}\partial_i
\left(\sqrt{-\gamma}
\gamma^{ij}\partial_j\right)~{\rm and}~\gamma={\mbox{det}}
\left(\gamma_{ij}\right)~.\nonumber
\end{equation}

We shall consider in this paper static and spherically symmetric
backgrounds;
this will simplify the analysis, but in principle one could carry
out the study for more general backgrounds also. We then take
\begin{equation}
ds^2=-a(r)dt^2+b(r)^{-1}dr^2+r^2d\Omega^2~,~~
d\Omega^2=d\theta^2+\sin^2\theta d\varphi^2~.\label{1.3}
\end{equation}
The zeroth order solution representing a spherical membrane can
be conveniently described by the following spherically symmetric
gauge choice
\begin{equation}
t=t(\tau)~,~~r=r(\tau)~,~~\theta=\rho~,~~\varphi=\sigma~,
\label{1.3'}\end{equation}
so that the induced metric on the world-volume becomes
\begin{equation}
\gamma_{\tau\tau}=-a\dot{t}^2+\dot r^2/b(r)~~,~~~
\gamma_{\rho\rho}=r^2~,~~
\gamma_{\sigma\sigma}=r^2\sin^2\rho~,
\end{equation}
where a dot denotes derivative with respect to tau.
Notice that the gauge choice (\ref{1.3'}) does not fix completely
the gauge for a
spherical membrane. There is still one un-used reparametrization,
which we
shall return to in a moment.

The effective Lagrangian of the system takes the following simple
form after
integrating over $\rho$ and $\sigma$
\begin{equation}
L=-4\pi T r^2\sqrt{a(r)\dot{t}^2-\dot r^2/b(r)}~.
\label{II.8}\end{equation}
The temporal component of the equations of motion (\ref{1.2}) is
given by
\begin{equation}
\partial_\tau\left({4\pi Tr^2a(r)\dot t\over\sqrt{a(r)\dot{t}^2-
\dot r^2/b(r)}}\right)=0~.
\label{1.4}
\end{equation}
The corresponding integral, $E$, has dimension of
$length^{-1}$\ =\ energy, when
using units where $c=\hbar=1,$
but keeping Newton's constant $G$ explicitly.
Let us define the proper time $\tau$ by
\begin{equation}
\dot t=\frac{E}{(4\pi T)^{1/3}a(r)}\equiv\frac{\tilde E}{a(r)}.
\label{II.10}\end{equation}
This definition fixes the remaining gauge freedom for
the spherical membrane.
Notice also that in a point-particle picture, the dimensionless
constant
$\tilde{E}$ would correspond to the energy per unit rest--mass
(at least in an asymptotically flat spacetime).
Introducing similarly the notation
$\tilde{T}=T/(4\pi T)^{1/3},$ the equation (\ref{1.4})
takes the form
\begin{equation}
\dot r^2={\tilde{E}^2 b\over a}\left(1-{(4\pi \tilde{T})^{2} r^4
\over\tilde E^2}a\right)\equiv\tilde E^2-~\tilde V^2,
\label{1.5}
\end{equation}
hence the effective potential can be read off \cite{MTW73}.
It is easy to check that the spatial components of the equations
of motion (\ref{1.2}) are consistent with (\ref{1.5}), and they
do not provide any further information. Thus the dynamics of the
spherical membrane is fully determined by Eq. (\ref{1.5}).

The condition of the existence of
a static spherical membrane $r=r_m$ (an equilibrium solution)
is given by $\dot{r}=0,\;(\tilde{V}^2)'=0$ at $r=r_m,$  where
the prime denotes derivative
with respect to $r,$ that is to say
\begin{equation}
4a(r_m)+r_ma'(r_m)=0~,
\label{1.6}
\end{equation}
while the condition for this membrane to be in {\it stable}
equilibrium (stable
with respect to radial perturbations; we shall consider arbitrary  
perturbations
in the next section) is
\begin{equation}
(\tilde V^2)''(r_m)=(4\pi \tilde{T})^{2}{b(r_m)\over a(r_m)}r_m^2
\left[-20a(r_m)+r_m^2a''(r_m)\right]>0~.
\label{1.7}
\end{equation}
It is remarkable that both conditions (\ref{1.6}), (\ref{1.7})
are actually independent of the function $b(r),$ at least as
long as we stay to the regime where $r$ is spacelike ($b(r)>0$).

It can now be easily shown by direct replacement into
Eq.\ (\ref{1.6}) that the simple assumption for the background
metric given by the Minkowski,
Schwarzschild or Reissner--Nordstr\"om metric, does not allow any
spherical membrane to be in (stable or unstable) equilibrium
outside the event horizon. This is easily understood from the
physical point of view: the membrane tension acts in the
direction of contracting the spherical membrane,
so in the absence of any other internal degrees of freedom
(which is the case for the Dirac membrane), a repulsive
gravitational field would be necessary to
support an equilibrium solution.  The simplest physically
interesting spacetime that admits solutions to Eq.\ (\ref{1.6})
is de Sitter spacetime, which in
static coordinates corresponds to $a(r)=b(r)=1-\Lambda r^2/3.$
The equilibrium solution is given by $r_m=\sqrt{2/\Lambda}$,
and it is unstable according to Eq.\ (\ref{1.7}).

However, as discussed in the Introduction, we are mostly
interested in black hole spacetimes.
In fact, let us take the Schwarzschild--de Sitter metric in
static coordinates. From now on we use Planck units
$(c=\hbar=G=1);$ we thus have
\begin{equation}
a(r)=b(r)=1-{2M\over r}-{\Lambda\over3}r^2~.
\label{II.14}\end{equation}
If (and only if)  $\Lambda M^2\equiv\tilde{\Lambda}\leq 1/9,$
there is a black hole event
horizon
$r_{EH}$ and a cosmological horizon $r_{CH}$
\begin{equation}
r_{EH}\leq r_{CH},
\end{equation}
and there is a unique solution $r_m$ to Eq.\ (\ref{1.6})
such that
\begin{equation}
r_{EH}\leq r_m\leq r_{CH}
\end{equation}
Explicit expressions for $r_{EH},\;r_m$ and $r_{CH}$
are given in Appendix\ A, and
Fig.\ \ref{fig1} shows the equilibrium membrane solution  $r_m$,
which always
lies
between the cosmological horizon $r_{CH}$ and the black hole event
horizon $r_{EH}.$
In Fig.\ \ref{fig2} we show  the effective potential
$\tilde V^2=(4\pi \tilde{T})^{2}a(r)r^4$,
and observe that the static spherical membrane sits at a local  
maximum of
the potential. We thus see that the equilibrium membrane solution
$r_m$ is
unstable
to radial perturbations. This can also be seen by
checking that the condition\ (\ref{1.7}) is not fulfilled for
any of the values of $r_m.$

\section{Covariant perturbations}
\setcounter{equation}{0}

To further understand the behavior of the membrane fluctuations
and stability we will now adapt the covariant approach of
Ref.\ \cite{LF94} (see also Refs.\ \cite{CG295,C93})
for strings, to the present case.

We have, for a membrane in four dimensions, only one degree of
freedom (as
opposed
to the two degrees of freedom for the string). This essentially
represents the fluctuations of the membrane perpendicular to the
hypersurface where it lies. The equation for this degree of
freedom, that we will denote as $\phi(\tau,\rho,\sigma),$
takes the form of a simple Klein--Gordon equation \cite{LF94}
(which had been obtained by in
Ref.\ \cite{L95} using the spherically symmetric gauge)
\begin{equation}
\Box_\gamma\phi+{\cal V}\phi=0~,\label{2.1}
\end{equation}
where
\begin{equation}
{\cal V}=\Omega_{ij}\Omega^{ij}-\gamma^{ij}
\partial_ix^\mu\partial_jx^\nu
R_{\mu\lambda\kappa\nu}n^\lambda n^\kappa~.\label{3.2}
\end{equation}
In the above equation, $R_{\mu\lambda\kappa\nu}$ is the Riemann  
tensor of the
background spacetime, while $\Omega_{ij}$ is
the second fundamental form
\begin{equation}
\Omega_{ij}=g_{\mu\nu}n^\mu x^\rho_{,i}\nabla_\rho x^\nu_{,j}~.
\label{3.3}\end{equation}
The normal vector $n^\mu$ is defined by
\begin{equation}
g_{\mu\nu}n^\mu x^\nu_{,i}=0,\;\;\;\;\;\;\;\;\;\;
g_{\mu\nu}n^\mu n^\nu=1,
\end{equation}
and it fulfills the completeness relation
\begin{equation}
g^{\mu\nu}=n^\mu n^\nu+\gamma^{ij} x^\mu_{,i} x^\nu_{,j}.
\end{equation}

Let us first consider a generic dynamical spherical membrane
in a background of the form (\ref{1.3}), as the zeroth
order solution. The normal vector is
\begin{equation}
n^\mu= \frac{\sqrt{b/a}}{\sqrt{\tilde{E}^2/a-\dot r ^2/b}}
\left(\dot r/b,\tilde{E},0,0\right)~,
\end{equation}
and then the components of the second fundamental form take the
following
explicit form for our background
metric\ (\ref{1.3}) \begin{eqnarray}
&&\Omega_{\tau\tau}=\frac{\tilde{E}\sqrt{b/a}}
{2a^2b\sqrt{\tilde{E}^2/a-\dot r
^2/b}}[2a^2\ddot{r}+\tilde{E}^2(ba'-ab')]+
\frac{\tilde{E}b'\sqrt{b/a}}{2b}\sqrt{\tilde{E}^2/a-
\dot r^2/b},\nonumber\\
&&\Omega_{\rho\rho}=\frac{-\tilde{E}r\sqrt{b/a}}
{\sqrt{\tilde{E}^2/a-\dot r
^2/b}}~,~~\;\;\;\;\;\;\;\;\;\;
\Omega_{\sigma\sigma}=\frac{-\tilde{E}r\sqrt{b/a}}
{\sqrt{\tilde{E}^2/a-\dot r
^2/b}}\sin^2\rho~,
\label{Omega}\end{eqnarray}
where primes denote $r$ derivatives.
These expressions are valid off--shell, that is,
we have not used the equation
of motion (\ref{1.4}). It is easy to check that the condition
$\gamma^{ij}\Omega_{ij}=0$ is equivalent to Eq. (\ref{1.4}),
as follows more
generally from the Gau\ss--Weingarten equation
\begin{equation}
D_{i}D_{j}x^\mu+\Gamma^\mu_{\kappa\lambda}\partial_i
{x^{\kappa}}
\partial_j{x^{\lambda}}=n^\mu\Omega_{ij}.
\label{III.8}\end{equation}
Here we are interested in the fluctuations around an on--shell
spherical Dirac
membrane. When using the equation of motion (\ref{1.5}),
the components (\ref{Omega}) reduce to
\begin{equation}
\Omega_{\tau\tau}=-2\tilde{E}r(4\pi
\tilde{T})\sqrt{b/a},\;\;\;\;\;\;\;\;
\Omega_{\rho\rho}\sin^2\rho=\Omega_{\sigma\sigma}=
\frac{-\tilde{E}\sqrt{b/a}}{r(4\pi \tilde{T})}\;\sin^2\rho.
\end{equation}
The first term of the potential (\ref{3.2}) then becomes
\begin{equation}
\Omega_{ij}\Omega^{ij}=\frac{6\tilde{E}^2 b}{ar^6(4\pi \tilde{T})}.
\end{equation}
To compute the second term of the potential, we need explicit
expressions for
the non-vanishing components of the
curvature tensor in the background metric (\ref{1.3})
\begin{eqnarray}
&&R_{rtrt}={1\over2}a''+{a'b'\over4b}-{(a')^2\over4a}~,~~
R_{r\theta r\theta}=-{rb'\over2b}~,~~
R_{r\varphi r\varphi}=-{rb'\over2b}\sin^2\theta~,\nonumber\\
&&R_{t\theta t\theta}={r\over2}ba'~,~~
R_{t\varphi t\varphi}={r\over2}ba'\sin^2\theta~,~~
R_{\theta\varphi\theta\varphi}=r^2(1-b)\sin^2\theta~,
\end{eqnarray}
We have now all the elements to write down explicitly
Eq.\ (\ref{2.1})
since
\begin{equation}
{\cal V}(r)=-\frac{b}{a}\left[\frac{a''}{2}+
\frac{a'}{r}+\frac{a'}{4ab}(ab'-ba')\right]
+\frac{b}{a}
\frac{\tilde{E}^2}{r^5(4\pi \tilde{T})^2}
\left[\frac{6}{r}-\frac{ab'-ba'}{ab}\right],
\end{equation}
and
\begin{equation}
\Box_\gamma=-\frac{1}{r^4(4\pi
\tilde{T})^{2}}\partial^2_\tau+\frac{1}{r^2\sin^2\rho}
\left( \sin\rho\ \partial_\rho(\sin\rho\partial_\rho)+
\partial^2_\sigma \right)
\end{equation}

Upon making the decomposition of the solution to
Eq.\ (\ref{2.1}) into spherical modes
\begin{equation}
\phi(\tau,\rho,\sigma)=\sum_{l,m}\phi_{l}(\tau)
Y_{lm}(\rho,\sigma)~,
\label{a15}
\end{equation}
where $Y_{lm}(\rho,\sigma)$ are the usual spherical harmonics,
we obtain
\begin{equation}
\ddot\phi_{l}+r^4 (4\pi \tilde{T})^{2}\left( {l(l+1)\over r^2}-
{\cal V}(r)\right)
\phi_{l}=0~.\label{2.2}
\end{equation}
This equation describes the fluctuations
around an on-shell dynamical spherical
membrane. For an equilibrium solution,
$r=r_m,$ it is equivalent to the
equation of motion of a harmonic oscillator with frequency
$\omega_l$
\begin{equation}
\omega^2_l=r_m^4 (4\pi \tilde{T})^{2}
\left({l(l+1)\over r_m^2}-{\cal V}(r_m)\right)
\label{III.16}\end{equation}
In the case of  Schwarzschild--de Sitter spacetime (SdS) we get
\begin{equation}
{\cal V}_{SdS}(r_m)= \frac{4r_m-9M}{r_m^3}~.
\label{III.17}\end{equation}
Since $0<{\cal V}_{SdS}(r_m)<4/r_m^2$ we see that
Eq.\ (\ref{2.2}) for $r=r_m,$ will
have real frequencies for $l>1$. Modes with $l=0,1$
generally possess imaginary frequencies, and are then
responsible for the instabilities we observed in the
previous section. In fact, a similar observation was already
made in Ref.\ \cite{GV91} in the study of dynamical spherical
membranes in de Sitter
spacetime. There, however, modes $l=0,1$ where not relevant
since they represented changes in the energy and linear
momentum of the membrane, and these where symmetries of the
background geometry. That is however not the case here, since
the black hole breaks Lorentz invariance and the instability
of the modes $l=0,1$ is physical. Notice, however, that for
$\tilde{\Lambda}\equiv \Lambda M^2\geq16/243$,
that is $r_m\leq 9M/2,$ the mode with $l=1$ is actually
stable (and, by the way, this is the case we are formally
interested in, c.f. the discussion in the Introduction,
since it corresponds to the membrane being relatively close
to the black hole event horizon, see Fig.\ \ref{fig1}), so
eventually
we are left with only the zero-mode $l=0$ being unstable.

\section{Higher order membrane}
\setcounter{equation}{0}

We have seen that a cosmological constant can provide the
necessary amount of repulsive gravitational field to ensure
the existence of spherical membranes in equilibrium in a
black hole background. Furthermore, by appropriately
choosing the dimensionless parameter
$\tilde{\Lambda}\equiv\Lambda M^2,$ the static
membrane can be at any position outside the black hole
event horizon. In particular, the static membrane can be at
the stretched horizon, only a few Planck lengths outside the
black hole event horizon. However, in the latter
case we must have a very large cosmological constant.
More generally,
the fact that $r_m$ rather
follows the cosmological horizon and only approaches the
black hole event horizon for large values of the cosmological
constant, see Fig.\ \ref{fig1}, lead
us to look for alternative ways of stabilizing a membrane.
In this section we shall consider extrinsic curvature
corrections to the membrane action \cite{CG95}.
They describe a membrane with finite thickness (the Dirac
membrane is infinitely thin).

We write the action in
curved spacetime as (where, in principle, $T$, $A$ and
$B$ are arbitrary constants)
\begin{equation}
I_{M}=\int d\tau d\rho d\sigma  \sqrt{-{\mbox{det}}
\left(\gamma_{ij}\right)}
\left[-T+A\left(\gamma^{ij}\Omega_{ij}\right)^2+
B\;\Omega^{ij}\Omega_{ij}\right]~,
\label{IV.1}\end{equation}
where the second fundamental form is defined in (\ref{3.3})
and where $A$ and $B$ have dimensions of $length^{-1}$.
In principle, we could also add a term proportional to the
scalar curvature of
the world-volume; it would be of the same order in
derivatives as the two terms
involving the second fundamental form\ \cite{BGMPU95}.
However, the purpose of the following
analysis is to show that equilibrium solutions can now be
obtained even without
the presence of a repulsive gravitational field. That is
to say, we will
eventually consider membranes in plain Schwarzschild
spacetime. In that case,
because the spacetime is Ricci--flat, the scalar curvature
of the world--volume is related to the two second fundamental
form terms, via the Gau\ss--Codazzi equation
\begin{equation}
R_{\gamma}=\left(\gamma^{ij}\Omega_{ij}\right)^2-
\Omega^{ij}\Omega_{ij},
\end{equation}
and therefore only two of these three terms are independent.

It is straightforward to derive the equations of motion from
the action (\ref{IV.1}). For a generic membrane, the trick
is to use the Gau\ss--Weingarten equation
(\ref{III.8}) to eliminate the dependence on the normal
vector \cite{gregory}.
The resulting equations, which are not particularly
enlightening
for the purposes of our paper,  contain up to four
derivatives in the world--volume coordinates.

Here we shall follow instead a somewhat simpler approach
directly adopted to
the spherical membranes.  We shall derive the effective
Lagrangian for a higher
order spherical membrane, i.e. the generalization of the
Dirac membrane effective Lagrangian (\ref{II.8})

Furthermore, in this section we choose to work in the spherically
symmetric rest gauge
\begin{equation}
t=\tau~,~~r=r(\tau)~,~~\theta=\rho~,~~\varphi=\sigma~,
\end{equation}
compare with Eqs.\ (\ref{1.3'}) and (\ref{II.10}).
We restrict ourselves to the case when $a=b$ in the
metric\ (\ref{1.3}). The
induced metric on the world--volume then takes the form
\begin{equation}
\gamma_{\tau\tau}=-a+\dot r^2/a~~,~~~
\gamma_{\rho\rho}=r^2~,~~
\gamma_{\sigma\sigma}=r^2\sin^2\rho~,
\end{equation}
and the components of the second fundamental form are
\begin{eqnarray}
&&\Omega_{\tau\tau}=\frac{\ddot{r}-aa'}{\sqrt{a-\dot{r}^2/a}}+
\frac{3a'}{2}\sqrt{a-\dot{r}^2/a},\nonumber\\
&&\Omega_{\rho\rho}=\frac{-ar}
{\sqrt{a-\dot r^2/a}}~,~~\;\;\;\;\;\;\;\;\;\;
\Omega_{\sigma\sigma}=\frac{-ar}{\sqrt{a-\dot r^2/a}}\sin^2\rho~.
\end{eqnarray}
It is now straightforward to derive the effective Lagrangian
corresponding to the action (\ref{IV.1})
\begin{eqnarray}
L=4\pi r^2\sqrt{a-\dot{r}^2/a}\;[-T&+&\frac{A+B}
{a-\dot{r}^2/a}
((\frac{\ddot{r}-aa'}{a-\dot{r}^2/a}+\frac{3a'}{2})^2+
\frac{2a^2}{r^2})
\nonumber\\
&+&\frac{A}{a-\dot{r}^2/a}
(\frac{4a}{r}(\frac{\ddot{r}-aa'}{a-\dot{r}^2/a}+\frac{3a'}{2})+
\frac{2a^2}{r^2})]
\end{eqnarray}
The variational principle leads to the equation
\begin{equation}
\partial^2_\tau\left( \frac{\delta L}{\delta \ddot{r}}\right)-
\partial_\tau\left( \frac{\delta L}{\delta \dot{r}}\right)+
 \frac{\delta L}{\delta {r}}=0,
\label{IV.7}\end{equation}
i.e., it contains up to four derivatives with
respect to time. In the first place, however, we will
look for a static
solution $r_m,$ corresponding to
$\partial_\tau r=\partial^2_\tau r=\partial^3_\tau
r=\partial^4_\tau r=0$. This condition reduces Eq.\ (IV.7) to
\begin{eqnarray}
T\left[2ra+{1\over2}r^2a'\right]\biggr\vert_{r=r_m}
&=&(A+B)\left[3aa'+{1\over2} r(a')^2 +{1\over2}r^2a'a''
-{r^2(a')^3\over8a}\right]\biggr\vert_{r=r_m}\nonumber\\
&+&A\left[5aa'+r(a')^2+2raa''\right]\biggr\vert_{r=r_m}
\label{IV.8}\end{eqnarray}

One can easily check that this condition reduces to
Eq.\ (\ref{1.6}) for $A=0=B$, i.e. when no higher order
terms are present.
Furthermore, this equation has no solutions at all for
$r>0$ in the Minkowski background,
i.e. when $a=1$. This last result is in agreement with that of
Ref.\ \cite{C94}. However, for generic $a(r)$ we will find
non-trivial solutions. This is a remarkable result:
As shown in \cite{C94}, the higher order terms cannot
support a static spherical membrane
in flat Minkowski spacetime, but we have now shown that
they can actually do it
in generic curved spacetimes, {\it even} in spacetimes
where the gravitational field is purely attractive.

In particular, for the simple
Schwarzschild metric (without cosmological constant!)
corresponding to $a(r)=1-2M/r,$ one does find  $r_m>2M$
solutions where a spherical membrane can be in
equilibrium:  In this case Eq.\ (\ref{IV.8}) reduces to
\begin{equation}
\alpha r^3_m(2r_m-3M)-2\beta Mr_m=
\frac{27M^3-26M^2r_m+6Mr^2_m}{r_m-2M}~.
\label{IV.9}\end{equation}
where
\begin{equation}
\alpha=\left(\frac{T}{A+B}\right)\quad\mbox{and}\quad
\beta=\left(\frac{A}{A+B}\right)~.
\end{equation}

Treating $T,$ $A$ and $B$ as arbitrary constants,
we can take $r_m$ at any
value outside the event horizon and we even have a
one-parameter family of ways
to fulfill Eq.\ (\ref{IV.9}). That is, we can for instance
fix the membrane tension
$T$ from the beginning but
still, choosing properly the otherwise arbitrary
constants $A$ and $B,$ pose this static membrane as close to
the event horizon as we wish, even on the stretched horizon.
In fact, for small $\epsilon$ we have
\begin{equation}
r_m=2M(1+\epsilon)~,
~~\epsilon\cong{1\over4(2\beta-4M^2\alpha-1)}~.
\end{equation}

We have thus established the existence of the static solution.
It remains open the question of the stability of this solution.
A complete and covariant discussion of the stability properties
of higher order membranes
involves the expansion of the action (\ref{IV.1})
up to second order in fluctuations
around a particular solution to the equations of motion.
In fact, that was how
Eqs.\ (\ref{2.1}), (\ref{3.2}) were obtained for the
Dirac membrane \cite{LF94}. Some
preliminary
steps of the analog computation for the higher order membrane
were taken in Ref.\ \cite{G93}, but the final equation
determining the fluctuations around an
arbitrary higher order membrane configuration embedded in an
arbitrary curved
spacetime has, to our knowledge, not yet been obtained in closed
form.

Here we will address only the question of the stability of
static spherical membranes (\ref{IV.8}) against zero-mode
fluctuations, that is, against radial fluctuations. We thus write
\begin{equation}
r=r_m+\phi(\tau),
\end{equation}
where $r_m$ is a solution to Eq.\ (\ref{IV.8}). We insert this
expression into the equation of motion (\ref{IV.7}) and keep
only terms linear in $\phi.$
After some
algebra, the resulting differential equation determining the radial
fluctuations takes the general form
\begin{equation}
\frac{d^4\phi}{d\tau^4}+F(r_m)\frac{d^2\phi}{d\tau^2}+
G(r_m)\phi=0~,
\label{IV.11}
\end{equation}
where $F(r_m)$ and $G(r_m)$ are complicated functions carrying the
information
about the static zeroth order solution and of the curved spacetime.
In the case
of the Schwarzschild black hole background, $a(r)=1-2M/r,$ they are
given by
\begin{equation}
F(r_m)=\left( \frac{r_m}{M}-1\right)\left[
3\alpha\left(1-\frac{2M}{r_m}\right)^2-
\frac{10Mr_m^2-45M^2r_m+54M^3}{r_m^5}\right],
\end{equation}
\begin{equation}
G(r_m)=\frac{-3\alpha M}{r_m^3}\left( \frac{r_m}{M}-1\right)
\left(1-\frac{2M}{r_m}\right)^2 +
\frac{7M^2r_m^2-27M^3r_m+27M^4}{r_m^8},
\end{equation}
and we have eliminated $\beta$ using Eq.\ (\ref{IV.9}).
The fluctuation equation (\ref{IV.11}) is solved by
\begin{equation}
\phi(\tau)=c_1 e^{d_1\tau}+c_2 e^{d_2\tau}+c_3 e^{d_3\tau}+
c_4 e^{d_4\tau},
\end{equation}
where $(c_1,c_2,c_3,c_4)$ are arbitrary constants, while
\begin{equation}
d_{(1,2,3,4)}=\pm\left( \frac{-F(r_m)\pm\sqrt{F^2(r_m)-4G(r_m)}}{2}
\right)^{1/2}.
\end{equation}
The necessary and sufficient condition for stability
is that $\phi(\tau)$ be oscillatory
($d_{(1,2,3,4)}$ purely imaginary). A necessary (but
not sufficient !) condition for this, is that both $F(r_m)>0$
{\it and} $G(r_m)>0.$ However, that condition leads to
\begin{equation}
10r^3_m-62Mr^2_m+126M^2r_m-81M^3<0,
\end{equation}
which can not be fulfilled outside the black hole event horizon.
Thus any static higher order membrane in the plain Schwarzschild
background is unstable with respect to radial fluctuations.

The main result of this section is that the higher order terms
in the membrane action can ensure the existence of static
spherical solutions in the background of a Schwarzschild black
hole. Thus we do not need the cosmological constant of
Secs.\ II--III, or any other kind of repulsive gravitational
field, for
that matter. However, the problem of instability (at least with
respect to
radial fluctuations) remained. In the next section
we shall discuss a way of by--passing these instability problems.

\section{Discussion}
\setcounter{equation}{0}

In this section we will show that it
is possible to avoid the instabilities
by considering a dynamical (contracting)
membrane instead of a static
equilibrium membrane. We will perform the
computations for a Dirac membrane
in the Schwarzschild--de Sitter background,
i.e. the model considered in
Secs.\ II--III. However, the results of
Sec.\ IV, concerning the
higher order membrane (especially the existence
of unstable static equilibrium
solutions), strongly indicate that the following
arguments can also be carried
through for the higher order membrane in the plain
Schwarzschild background, that is to say,
the presence of the cosmological constant does not seem to be   
essential.

Let us return to the potential (\ref{1.5}) in the case of the
Schwarzschild--de Sitter spacetime. It is apparent from 

Fig.\ \ref{fig2}
that there are the following qualitatively different kinds of
``orbits'' for spherical membranes

If $\tilde E^2>\tilde V^2(r_m)$ membranes are always ``over''
the potential barrier, and either expand for ever or contract
towards
the black hole, ending trapped by the singularity at $r=0$.

If $\tilde E^2<\tilde V^2(r_m)$ membranes might bounce if they
are expanding and are ``inside'' the potential well, or if they
are contracting but are ``outside'' the potential well.

We are particularly interested in the intermediate case, i.e.
$\tilde E^2=\tilde V^2(r_m),$ of which we have analyzed the
stability properties of the static equilibrium configuration
$r=r_m$ in Sec.\ II. However, for $\tilde E^2=\tilde V^2(r_m),
$ there is also a contracting solution as well as an expanding
solution. For the static solution we have:
\begin{equation}
\tilde E^2=(4\pi\tilde{T})^2 r^4_m
\left( 1-\frac{2M}{r_m}-\frac{\Lambda}{3}r^2_m\right),
\label{V.1}\end{equation}
\begin{equation}
2-\frac{3M}{r_m}-\Lambda r^2_m=0.
\label{V.2}\end{equation}
Now consider a contracting solution with "energy" given by
(\ref{V.1}). We are interested in solutions contracting
from a finite coordinate distance $r_0$ (where $r_m<r_0<r_{CH}$)
towards $r_m.$ The {\it proper}
time to reach the summit of the potential barrier ($r=r_m$)
happens to be logarithmically divergent, and can be derived
as follows from Eq.\ (\ref{1.5})
\begin{equation}
\int_{\tau_0}^{\tau_m}d\tau=-\int_{r_0}^{r_m}{dr\over
\sqrt{\tilde E^2-\tilde V^2(r)}}=
-{1\over (4\pi\tilde{T})M}\int_{x_0}^{x_m}{dx\over
(x-x_m)\sqrt{f(x)}}~,
\label{V.3}\end{equation}
where
\begin{equation}
f(x)={1\over3}\left\{\tilde\Lambda x^4+2\tilde\Lambda x^3x_m+
3(\tilde\Lambda x_m^2-1)x^2+2(2\tilde\Lambda x_m^3-3x_m+3)x+
5\tilde\Lambda x^4_m-9x^2_m+12x_m\right\},
\end{equation}
and we have taken the dimensionless variables $x=r/M$ and
$\tilde\Lambda=\Lambda M^2$. One can thus show that the proper
time distance becomes infinite. In fact,
\begin{equation}
\tau_m-\tau_0=-{1\over(4\pi\tilde T)M}{1\over\sqrt{f(x_m)}}
\ln(x-x_m)\biggr\vert_{x_0}^{x_m}+...
\label{V.5}\end{equation}
where
\begin{equation}
f(x_m)=x_m\left(6-6x_m+5x_m^3\tilde\Lambda\right)~,
\end{equation}
and the ellipsis stands for finite terms as $x\to x_m$.

We have stressed that this infinity occurs in the proper time
description. This is clearly different from the infinite
``Schwarzschild'' time it takes a particle to reach the event
horizon (while it only takes a finite proper time, even to reach
the singularity). In fact, the above logarithmic divergence
arises at a coordinate radius $r_m>r_{EH}$. This is a particular
property of the Schwarzschild--de Sitter spacetime background,
that generates a maximum in the effective potential of the
membrane, and this does {\it not} occur for the Dirac membrane
in the plain Schwarzschild
geometry.

On the other hand, the above discussion applies quite generally
to any potential barrier possessing a maximum. The infinite
time to reach the top of the potential is well--known in
classical mechanics, for instance, an ordinary pendulum
experiences the same process when going towards the vertical
(unstable) position with precisely the energy corresponding to
being in the static, unstable equilibrium position.
The results of Sec.\ IV show that the dynamics of the higher
order membrane in the plain Schwarzschild background is also
governed by some kind of finite potential barrier, thus also
in that case we can expect to find the above mentioned type of  
solution.

We now consider fluctuations around
the contracting Dirac membrane in the
Schwarzschild--de Sitter background,
using equation (\ref{2.2}). In general,
in the Schwarzschild--de Sitter spacetime
\begin{equation}
{\cal V}(r)=\Lambda+\frac{6\tilde{E}^2}{(4\pi\tilde{T})^2 r^6},
\end{equation}
so that for the contracting membrane with $\tilde{E}$ given by
Eq.\ (\ref{V.1})
\begin{equation}
{\cal V}(r)=\Lambda+\frac{6r^4_m a(r_m)}{r^6},\;\;\;\;\;\;\;r_m<r<r_0
\end{equation}
where $a(r)$ is given by (\ref{II.14}) and $r_m$ is the solution
of\ (\ref{V.2}).
The "frequency" in equation (\ref{2.2}) now also depends on $r$
\begin{equation}
\omega_l^2(r)=r^4 (4\pi\tilde{T})^2\left(
\frac{l(l+1)}{r^2}-\Lambda-\frac{6r^4_m a(r_m)}{r^6}\right).
\label{V.9}\end{equation}
By careful (and partly numerical) study of the formulas for
$r_m$ and $r_{CH}$ (See Appendix\ A), one finds that:
\begin{eqnarray}
&\omega_0^2(r)<0&\nonumber\\
&\omega_1^2(r)>0&\;\;\;\;\;\Leftrightarrow
\;\;\;\;\tilde{\Lambda}>0.07737...\\
&\omega_l^2(r)>0&\;\;\;\;\;\;\;\;\;\;\;\;\;\;
\mbox{for all}\;\;l\geq 2\nonumber
\end{eqnarray}
and these results hold for all $r_m\leq r\leq r_0.$ Notice that the  
condition
for $\omega_1^2$ being positive is somewhat more restrictive than  
what was
obtained at the end of Sec.\ III. The reason is that we now consider
$\omega_1^2$ for arbitrary $r,$ not only for $r=r_m.$

We are particularly interested in the
case when the membrane approaches a
position close to the Schwarzschild
event horizon. This is the case when
$\Lambda M^2\equiv\tilde{\Lambda}$ is
close to the maximal value
$\tilde{\Lambda}_{\mbox{max}}=1/9,$ c.f.
the discussion at the end of
Sec.\ III. From the above results follow
that for such contracting
membranes, {\it all} $\;l\geq 1$--modes are
stable everywhere (that is, for all
$r$) during the
contraction. At first sight it seems, on
the other hand, that the zero-mode
is always unstable. However, for a
contracting spherical membrane, the
zero-mode does not represent a real
physical fluctuation. The zero-mode
merely represents a time-translation,
and therefore should be eliminated
from the spectrum. We thus conclude that
{\it all} physical fluctuations
around the contracting membrane are stable.

To actually compute the classical and quantum
spectrum of physical fluctuations
($l\geq 1),$ we must solve the equation
\begin{equation}
\ddot{\phi}_l+\omega_l^2(r) {\phi}_l=0,
\label{V.11}\end{equation}
where $\omega_l^2(r)$ is given by (\ref{V.9}), and $r=r(\tau)$
is obtained by inverting (\ref{V.3}). Eq.\ (\ref{V.3}) can
be solved explicitly in terms of
Jacobi Elliptic Functions \cite{AS65}, and Eq.\ (\ref{V.11})
is then of (generalized) Lam\'{e}-type. However, we shall not
attempt here to obtain the exact solution for $\phi(\tau)$ in
closed form. Instead we will use the result that the membrane
takes infinite proper time to reach the summit of the
potential, c.f. Eq.\ (\ref{V.5}). Physically this means
that after some finite time, the membrane
is effectively always very close to the top of the barrier.
We can thus approximate $\omega_l^2(r)$ by a constant which
is essentially $\omega_l^2(r_m),$ as given by
Eqs.\ (\ref{III.16}) and (\ref{III.17}). We are then
left with an ordinary harmonic
oscillator equation, which can be easily solved. In fact, in
Ref.\ \cite{L95} it was assumed that this equilibrium point 
existed and from the quantization of the harmonic oscillations
a discrete spectrum of energies for the membrane obtained.
The entropy associated to this discrete energy levels, in the
thermal bath of the background black hole, can be computed and
one obtains a leading term proportional to the two--surface of
the membrane plus a non--leading logarithmic term.

In conclusion, we have learned that in a curved background, as
opposite to what happens in flat space,
spherical membranes can exist in equilibrium.
Higher order membranes provide enough
arbitrariness to locate the membrane as close to the event
horizon as we want, even in the plain Schwarzschild background.
The issue of the stability of spherical
membranes is delicate. We identified modes $l=0$ and sometimes
also $l=1$ as responsible for the instabilities. We discussed,
however, a certain range of validity of the perturbative
description and eventual quantization of the fluctuations.
Further study of the non--radial fluctuations of higher order
membranes is needed, as well as of the situation
in more general curved backgrounds (for example, to include
rotation) to see if a ``true'' stability mechanism can be found.

\begin{acknowledgments}
A.L.L. was supported by NSERC (National
Sciences and Engineering Research Council of Canada), while
C.O.L was supported by the NSF grant PHY-95-07719 and by research
founds of the University of Utah.
\end{acknowledgments}
\newpage

\appendix
\section{Horizons and Static Membrane in Schwarzschild--de Sitter}
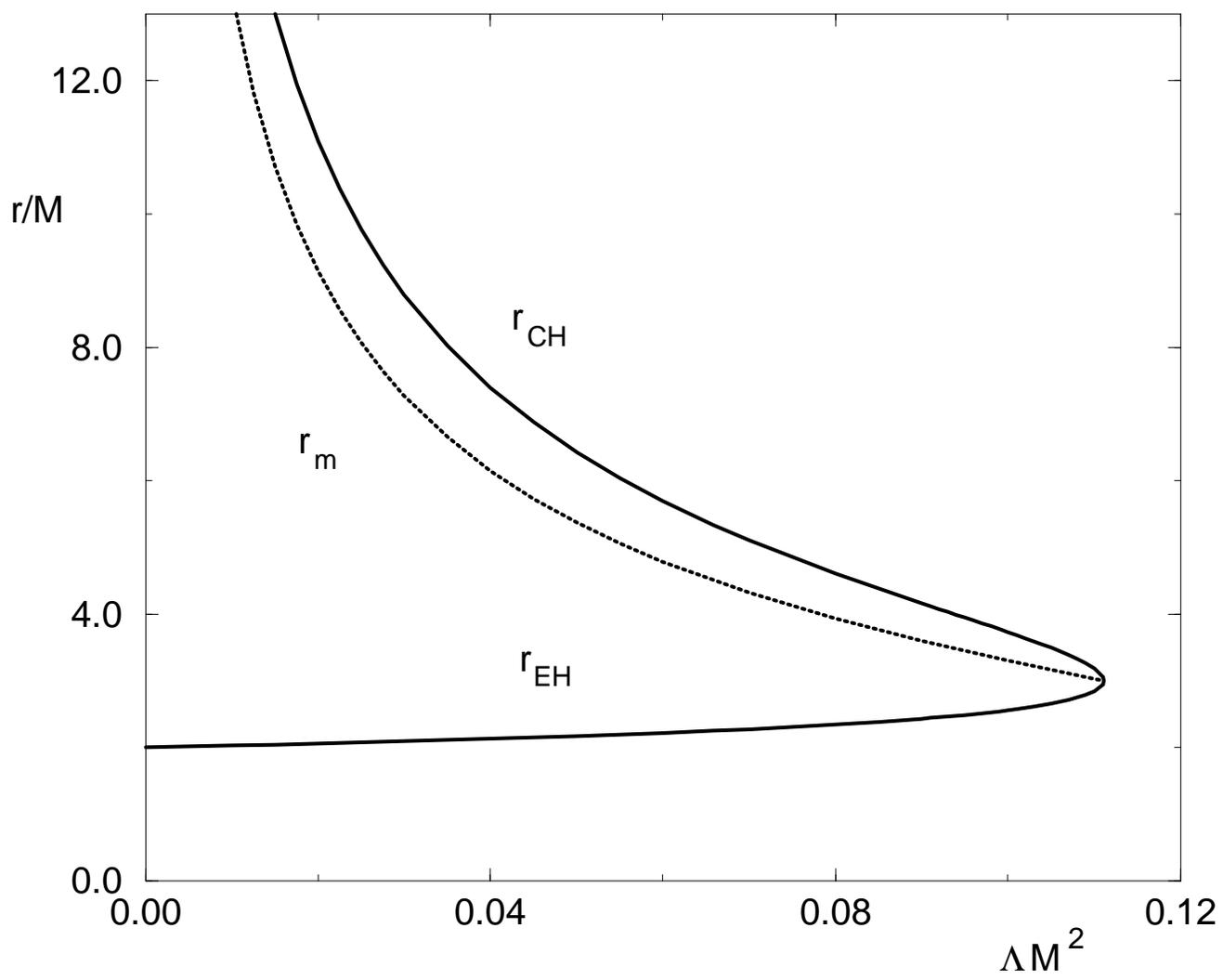
\begin{figure}
\caption{We observe that the radius at which a spherical membrane
is at equilibrium $r_m,$ in Schwarzschild--de Sitter space, lies  
between the
cosmological horizon
$r_{CH}$
and the black hole event horizon $r_{EH}$. It rather folows the
cosmological horizon and only approaches the event horizon for
relatively large values of the cosmological constant. The three
radii merge together at $r=3M$ and
$\tilde{\Lambda}=\Lambda M^2=1/9$. Higher values of the
cosmological constant lead to a naked singularity,
see Appendix\ A.}
\label{fig1}
\end{figure}

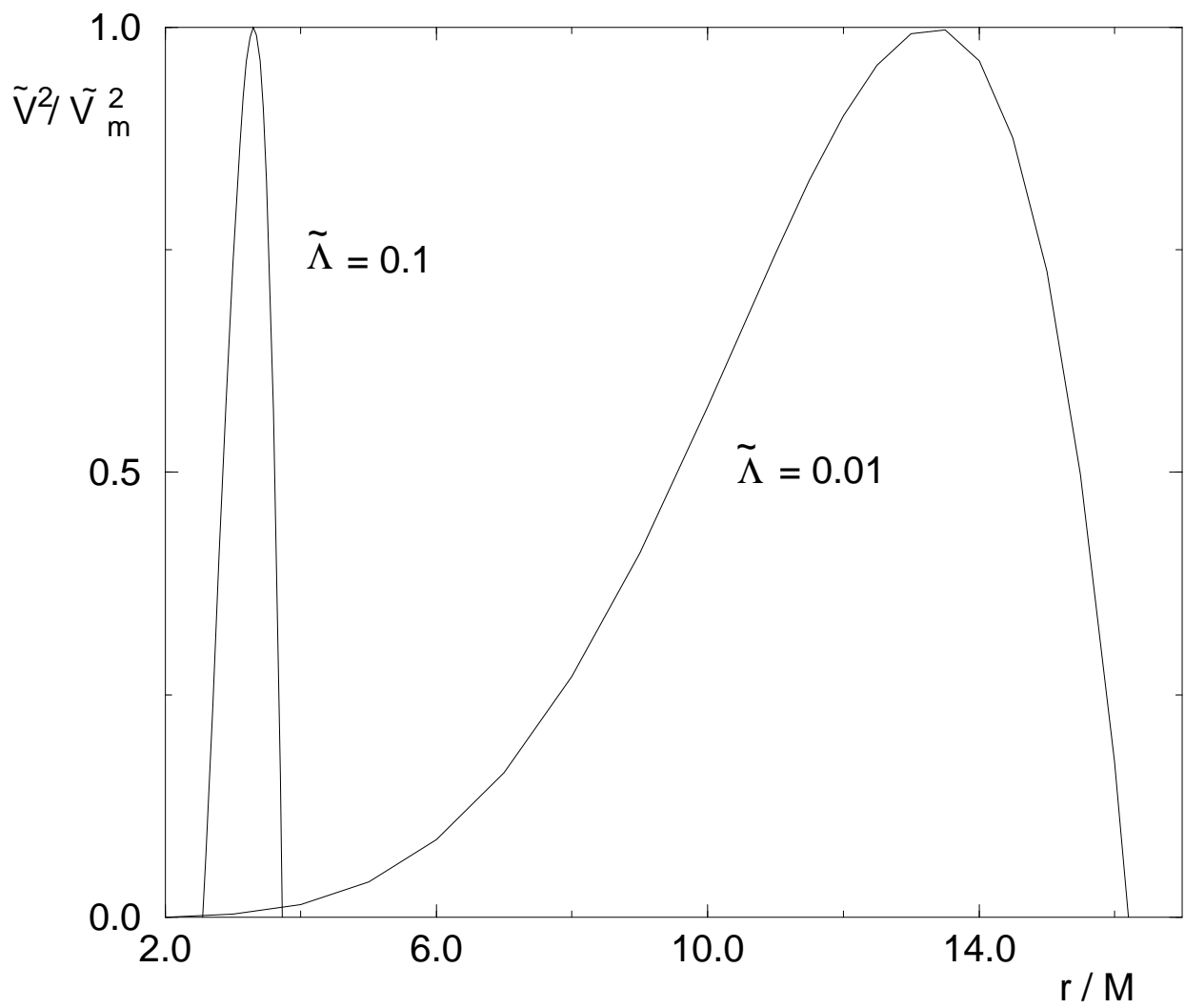
\begin{figure}
\caption{The effective potential (normalized to its maximum
value $V_m$) felt by a spherical membrane in
Schwarzschild--de Sitter background for two different values of
the cosmological constant $\tilde\Lambda=\Lambda M^2$. For the
values of $\tilde\Lambda=0.1,\;0.01$ shown in the figure, the
coordinates of the maximum are $r_m/M=3.305,\ 13.32,$ respectively.
As $\tilde\Lambda$ decreases, $r_m$ increases and the potential
moves to the right. In the other
extreme, $\tilde\Lambda=1/9$, we have $r_m=3M$ and $V_m=0$.}
\label{fig2}
\end{figure}

\setcounter{equation}{0}
Provided that $M>0$ and $\Lambda\geq 0,$ the equation
\begin{equation}
1-\frac{2M}{r_H}-\frac{\Lambda}{3}r_H^2=0
\end{equation}
has real positive solutions if and only if
$\Lambda M^2\equiv\tilde{\Lambda}\leq 1/9.$ In that case,
the cosmological horizon $r_{CH}$ and the black hole event horizon
$r_{EH}$ are
given by:
\begin{equation}
\frac{r_{CH}}{M}=Z^{-1}+{Z\over\tilde\Lambda}~,
\end{equation}
\begin{equation}
\frac{r_{EH}}{M}=-{1\over2}\left[{1-i\sqrt{3}\over Z}+
{(1+i\sqrt{3})Z\over\tilde\Lambda}\right]~,
\end{equation}
where
\begin{equation}
Z=\left(i\sqrt{\tilde\Lambda^3(1-9\tilde\Lambda)}-
3\tilde\Lambda^2
\right)^{1/3}.
\end{equation}
The radial coordinate of a static Dirac membrane, solution of
Eq.\ (\ref{1.6}),
\begin{equation}
2-3{M\over r_m}-\Lambda r_m^2=0
\end{equation}
is explicitly given by
\begin{equation}
\frac{r_{m}}{M}={2\over W}+{W\over3\tilde\Lambda}~,
\end{equation}
where
\begin{equation}
W=2^{-1/3}\left(3i\sqrt{3\tilde\Lambda^3(32-243\tilde\Lambda)}-
81\tilde\Lambda^2\right)^{1/3}.
\end{equation}
In the limiting cases we have
\begin{equation}
\tilde{\Lambda}=0\; :
\;\;\;\;\;\;r_{EH}=2M,\;\;\;\;\;\;r_{m}=r_{CH}=\infty,
\end{equation}
\begin{equation}
\tilde{\Lambda}=1/9\; : \;\;\;\;\;\;\;\;\;\;\;  
r_{EH}=r_{m}=r_{CH}=3M,
\end{equation}
while in general
\begin{equation}
r_{EH}\leq r_{m}\leq  r_{CH}.
\end{equation}
These results are illustrated in Fig.\ \ref{fig1}.

\end{document}